\documentclass[final]{aipproc}

\layoutstyle{6x9}

\begin{document}

\title{Generalized parton distributions of nuclei}


\author{V. Guzey}{
  address={Theory Center, Thomas Jefferson National Accelerator Facility, 
Newport News, VA 23606, USA}, email={vguzey@jlab.org}}

\begin{abstract}

We review recent theoretical results on generalized parton distributions (GPDs) of nuclei, emphasizing the following three roles of nuclear GPDs: (i) complementarity to free proton
GPDs, (ii) the enhancement of traditional nuclear effects such as nuclear binding, 
EMC effect, nuclear shadowing, and (iii) an access to novel nuclear effects such as 
medium modifications of bound nucleons.

\end{abstract}

\maketitle


\section{Introduction}

The theoretical and experimental interest to generalized parton distributions 
(GPDs) of hadrons is based on the facts that GPDs:
\begin{itemize}
\item[(i)] provide a rigorous theoretical framework for the 
description of hard exclusive processes (deeply virtual Compton scattering, deep
exclusive electroproduction of mesons, etc.) that is based on the QCD factorization
theorem,
\item[(ii)] extend the traditional one-dimensional (longitudinal) picture of hadrons to the full three-dimensional one,
\item[(iii)] help to understand the spin structure of hadrons in terms of underlying
quarks and gluons.
\end{itemize}
While the above points are valid for any hadronic target, nuclear GPDs 
are interesting and important in their own right:
\begin{itemize}
\item[(i)]
Nuclear GPDs provide the information on nucleon GPDs which is complimentary to 
that of the GPDs of the free proton,
\item[(ii)]
Traditional nuclear effects (nuclear binding, EMC effect, nuclear shadowing) are enhanced
in nuclear GPDs compared to the usual nuclear parton distributions,
\item[(iii)]
 Nuclear GPDs give an access to such novel (or at least not well-established) nuclear effects as the presence of non-nucleon degrees of freedom and 
medium modifications of the GPDs of the bound nucleon.
\end{itemize}
Below we review recent theoretical results on each of the three aspects of nuclear 
GPDs that we have just emphasized.

\section{Complimentary role of nuclear GPDs}

Since nuclei consist of protons and neutrons, models of nuclear GPDs require
proton and neutron GPDs as input~\cite{Guzey:2003jh,Kirchner:2003wt,Guzey:2008th}.
Therefore, measurements of deeply virtual Compton scattering (DVCS) with nuclear targets
in the coherent (the target nucleus stays intact) and incoherent (the target nucleus breaks up)
regimes provide additional constraints on the nucleon GPDs, which are complimentary to the
information extracted from DVCS on a free proton. The examples include
the completed and analyzed Hermes measurement of DVCS on a range of nuclear targets 
($^4$He, $^{14}$N, $^{20}$Ne, $^{84}$Kr and $^{131}$Xe)~\cite{Ye:2008zz} and
 the future high-precision measurement of DVCS on $^4$He at 
Jefferson Lab~\cite{He4-JLab}.

\section{Nuclear GPDs and conventional nuclear effects}

\subsection{Nuclear binding and off-diagonal EMC effect}

The EMC effect is the observation made by the European Muon Collaboration (EMC) in 1983
that the nuclear structure function measured in inclusive deep 
inelastic scattering (DIS) is smaller than the sum of the free nucleon structure functions, 
$F_{2A}(x,Q^2) < A F_{2N}(x,Q^2)$, for $0.2 < x < 0.7$, where $A$ is the number of nucleons.
After 25 years of investigations, there is no universally accepted explanation of the EMC effect. 
In particular, the conventional nuclear binding mechanism explains only a small fraction of the observed
EMC suppression.
Hence, one hopes that the studies of the EMC effect in DVCS with nuclear targets (in off-diagonal 
kinematics) may shed some light on the origin of the EMC effect.

The traditional nuclear binding effect in nuclear GPDs can be taken into account
using the impulse (convolution) approximation~\cite{Scopetta:2004kj,Guzey:2005ba,Scopetta:2009sn}:
\begin{equation}
H_A^q(x,\xi,t)=\sum_N \int_x^1 \frac{dz}{z} \,h_A^N(z,\xi,t)\, H_N^q\left(\frac{x}{z},\frac{\xi}{z},t \right) \,,
\label{eq:convol}
\end{equation}
where $H_A^q$ is the quark GPD ($q$ is the quark flavor) in a nucleus; $H_N^q$ is the quark 
GPD in the nucleon; $h_A^N$ is the off-diagonal light-cone distribution of nucleons in the 
target nucleus. The involved GPDs depend on the standard variables $x$, $\xi$ and $t$; 
$z$ is the light-cone fraction of the interacting nucleon.

The calculation for $^3$He using $h_A^N$ obtained with the off-diagonal spectral function and realistic
nucleon-nucleon potential demonstrates that the ''off-diagonal EMC effect'', i.e.~the deviation of
the ratio $H_{^3 {\rm He}}^q/(2 H^q_p+H^q_n)$ from unity, is larger than in the forward limit, strongly 
depends on the quark flavor and has the form, which is very different from the conventional EMC effect~\cite{Scopetta:2009sn}. This is a very important result for the interpretation of the nuclear 
DVCS data, which stresses that nuclear GPDs have amplified conventional nuclear effects that should not 
be misinterpreted as new, exotic nuclear effects.

One should also mention important results on the theory of GPDs of the 
deuteron~\cite{Berger:2001zb} and on the studies of the GPDs of the deuteron in the impulse
approximation on the light-cone~\cite{Cano:2003ju}.

\subsection{Nuclear GPDs and nuclear shadowing}

In inclusive DIS with nuclear targets, nuclear shadowing is the depletion of the nuclear
structure function compared to the sum of the nucleon structure functions, 
$F_{2A}(x,Q^2) < A F_{2N}(x,Q^2)$, for $x < 0.1$. This also translates into
nuclear shadowing 
for nuclear
quark and gluon parton distributions (PDFs). The origin of nuclear shadowing
is fairly well-understood. In the target rest frame, it originates from
multiple simultaneous interactions of the projectile with all nucleons of the target.
Observing that each interaction with the target nucleons has a diffractive character and
using the QCD factorization theorem coupled with the QCD analysis of the HERA data on
inclusive diffraction in electron-proton DIS, one arrives at the so-called leading twist theory of nuclear shadowing for nuclear PDFs.

\begin{figure}
\includegraphics[height=.24\textheight]{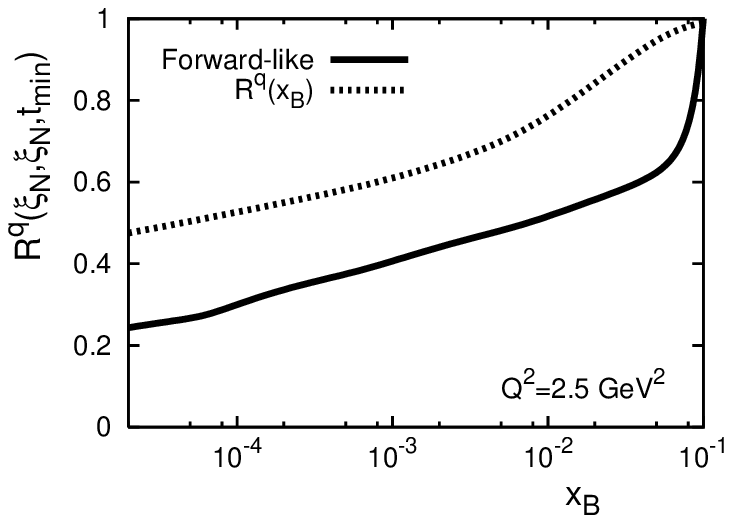}
\includegraphics[height=.24\textheight]{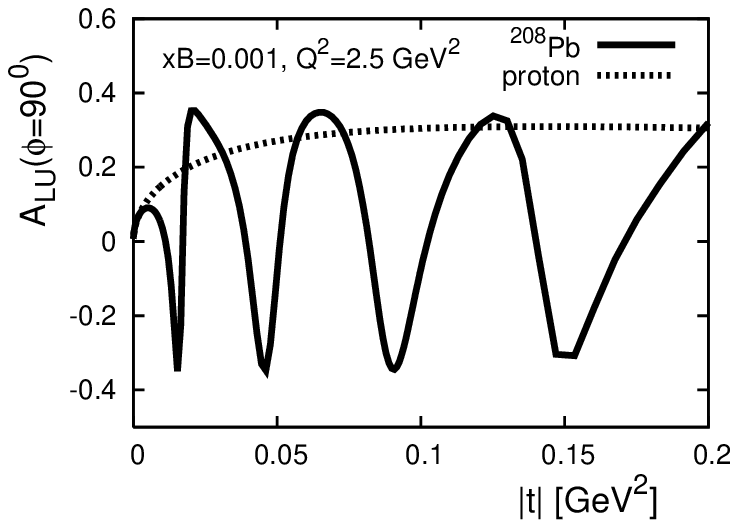}
\caption{Left panel: Nuclear shadowing for nuclear GPDs of $^{208}$Pb.
The ratio $R$ of Eq.~(\ref{eq:R_shadowing}) as a function of Bjorken $x$ (solid curve).
For comparison, nuclear shadowing for usual nuclear quark PDFs is also given (dotted curve).
Right panel: The DVCS beam-spin asymmetry as a function of $|t|$. The solid curve is for
$^{208}$Pb, the dotted curve is for the free proton.
}
\label{fig:guzey_shadowing}
\end{figure}

The leading twist theory of nuclear shadowing can be generalized to DVCS and to nuclear GPDs~\cite{Goeke:2009tu}. In the resulting model, nuclear shadowing in nuclear GPDs is
modelled using the GPDs of the Pomeron.
The left panel of Fig.~\ref{fig:guzey_shadowing} presents an example of the resulting 
predictions for nuclear shadowing for nuclear GPDs as the ratio $R(\xi,\xi,t_{\rm min})$,
\begin{equation}
R(\xi,\xi,t_{\rm min})=\frac{H_A^q(\xi,\xi,t)}{F_A(t) \sum_N H_N^q(\xi,\xi,t)} \,,
\label{eq:R_shadowing}
\end{equation}
where $F_A(t)$ is the nuclear form factor. In the absence of nuclear shadowing, 
$R(\xi,\xi,t_{\rm min})=1$. As one can see from
the figure, the solid curve representing $R(\xi,\xi,t_{\rm min})$ dips significantly below
unity at small Bjorken $x$. This deviation from unity
(nuclear shadowing) is larger than that
for usual nuclear PDFs (dotted curve). 

One can also examine how nuclear shadowing will affect
nuclear DVCS observables. The right panel of Fig.~\ref{fig:guzey_shadowing} presents 
predictions~\cite{Goeke:2009tu} for the nuclear DVCS beam-spin asymmetry, $A_{\rm LU}$,
 as a function of the momentum transfer $|t|$ (solid curve). For comparison, $A_{\rm LU}$
for the proton is given by the dotted curve. It is the effect of nuclear shadowing
that causes the dramatic oscillations of $A_{\rm LU}$ for nuclear targets.

\section{Medium modifications of bound nucleon GPDs}

Properties of the bound nucleons are expected to be modified by the nuclear medium.
The examples include the modifications of the structure functions (EMC effect) and 
elastic form factors (recoil polarization in quasi-elastic scattering on $^4$He), 
suppression of the axial coupling constant, and many more. Therefore, it is natural to 
assume that GPDs of the bound nucleon are also modified in nuclei.
Such modifications can be probed in incoherent DVCS with nuclear target, 
and, in particular, by the future Jefferson Lab experiment with 
$^4$He~\cite{He4-JLab}.

Recalling the connection between GPDs and elastic form factors, one can propose a
simple model that the bound nucleon GPDs are modified in proportion to the corresponding
elastic form factors~\cite{Guzey:2008fe}. For the medium modified elastic form factors,
the results of the Quark Meson-Coupling model were used, which are consistent with the
data on recoil polarization in quasi-elastic scattering on $^4$He, see details in~\cite{Guzey:2008fe}.

\begin{figure}
\includegraphics[height=.24\textheight]{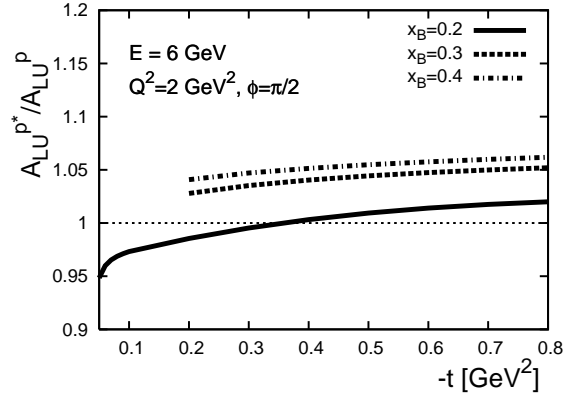}
\caption{The ratio of the beam-spin asymmetries of the proton
bound in $^4$He to the free one~\cite{Guzey:2008fe} in the kinematics of the future Jefferson Lab experiment~\cite{He4-JLab}.
}
\label{fig:gpds_mm}
\end{figure}

The resulting GPDs of the bound nucleon can be used to make predictions for various observables measured in incoherent nuclear DVCS.
Figure~\ref{fig:gpds_mm} presents the ratio of the beam-spin asymmetries of the proton
bound in $^4$He to the free one in the kinematics of the future Jefferson Lab experiment~\cite{He4-JLab}.

Note that the predictions of Ref.~\cite{Guzey:2008fe} presented in Fig.~\ref{fig:gpds_mm} are different
from the only other existing prediction for the medium modifications of the bound nucleon GPDs~\cite{Liuti:2005gi}.

\bibliographystyle{aipproc}   

\end{document}